\documentclass[aps,prl,twocolumn,showpacs,superscriptaddress,letterpaper,amsmath,amssymb]{revtex4} 

\usepackage{graphicx}  % needed for figures
\usepackage{dcolumn}   % needed for some tables
\usepackage{bm}        % for math
\usepackage{amssymb}   % for math
\usepackage{feynmf}%%%{feynmp}
\usepackage{slashed}
\def\gsim{\lower0.5ex\hbox{$\:\buildrel >\over\sim\:$}}
\def\lsim{\lower0.5ex\hbox{$\:\buildrel <\over\sim\:$}}

\newcommand{\be}{\begin{equation}}
\newcommand{\ee}{\end{equation}}
\newcommand{\bea}{\begin{eqnarray}}
\newcommand{\eea}{\end{eqnarray}}

\newcommand{\nbox}{{\,\lower0.9pt\vbox{\hrule \hbox{\vrule height 0.2 cm
\hskip 0.2 cm \vrule height 0.2 cm}\hrule}\,}}

\begin{document}

\thispagestyle{empty}
\vspace*{-3.5cm}

\vspace{0.5in}

%\begin{flushright}
%\today\\
%\end{flushright}
%\vspace{0.5in}
\title{Limits on Four-Top Production from the ATLAS Same-sign Top-quark Search}

\begin{center}
\begin{abstract}
We repurpose the recent ATLAS search for same-sign top quarks in data
with 1.0 fb$^{-1}$ in the context of a search for production of four
top quarks.  Using the null results of that search, we place limits on the four-top-quark production cross section of
about 1 pb.  These limits are  larger than the expected Standard Model
rate for four-top-quark production, but are already
strong enough to place interesting constraints on models which enhance that rate.  We interpret these results in the
context of models in which the right-handed top quark is composite and find limits on the compositeness scale of
about 700 GeV.
\end{abstract}
\end{center}

\author{Ning Zhou}
\author{Daniel Whiteson}
\author{Tim M.P. Tait}
\affiliation{Department of Physics \& Astronomy, University of California, Irvine, CA 92697}
\preprint{UCI-HEP-TR-2012-03}
\maketitle

The top quark, as the heaviest fermion of the Standard Model (SM), and the only one to have a mass of
``natural" size at the electroweak scale, is a natural laboratory to explore physics beyond the Standard Model.
In particular, while production and decay of top quarks seems to be roughly in line with SM expectations
\cite{Lannon:2012fp} (however, see \cite{Kamenik:2011wt}), 
many open questions remain.  For example, many
models such as top-color \cite{Hill:1991at}
and top-flavor \cite{Chivukula:1994mn}
posit a new interaction unique to the top quark, which ultimately drives electroweak symmetry-breaking.
Such an interaction would most naturally reveal itself through
multi-top-quark production
\cite{Georgi:1994ha,Lillie:2007hd}, 
a process requiring at least $4 m_t \sim$~TeV partonic energies that are just now
becoming accessible to the Large Hadron Collider.  Even if the
interactions of the top quark itself
are more prosaic, there may still be enhancements of the four-top-quark production rate arising from
gluino decays \cite{Kane:2011zd}
(which also lead to additional sources of missing energy in the events) or 
exotic colored scalar \cite{Gerbush:2007fe} decays.

One particularly fascinating vision for physics which contributes to
multi-top-quark  production are models
where the top quark is composite \cite{Georgi:1994ha,Lillie:2007hd,Pomarol:2008bh,Kumar:2009vs,Nomura:2009tw}.
Such models fit easily within the paradigm of strongly coupled theories of electroweak symmetry-breaking
\cite{ArkaniHamed:2000ds}
dual to Randall-Sundrum (RS) models of an extra dimension \cite{Randall:1999ee},
but also are interesting in their own right, as complements to searches for compositeness of the light quarks
and/or leptons.  Without a detailed picture for the UV physics, it is difficult to make firm predictions for
how a composite top quark would manifest at the LHC.  However, generic features such
as a four-top-quark contact interaction or the presence of a color-octet vector particle with strong
coupling to the top quark (reminiscent of the RS Kaluza-Klein gluon \cite{Agashe:2006hk}) are expected to appear in
wide classes of models \cite{Lillie:2007hd}.  Limits from the Tevatron derived from
the rate of $t \bar{t}$ pair production are rather weak \cite{Lillie:2007hd,Kumar:2009vs}, placing essentially
no useful bound on the compositeness scale if there are modest cancellations in the leading modifications
to the gluon-top-anti-top vertex.

Given the rich physics potential of multi-top-quark events, in this letter we perform an analysis of
LHC data to search for four-top-quark production.  We rely on the
cases in which either both top quarks or both
anti-top-quarks decay leptonically, leading to a same-sign dilepton signature with small expected SM background.
Our results are derived from the recent ATLAS same-sign dilepton
(plus jets and missing transverse momentum) search~\cite{atlasss}
based on 1.0 fb$^{-1}$ of integrated luminosity, which was originally
designed to search for same-sign top-quark pair production and fourth generation quarks.
For an alternate proposal to search for events with four top quarks in
early LHC running, see Ref.~\cite{Gregoire:2011ka}.
We interpret our results in the context of four-top-quark production through
a color-octet ($\rho_o$) or color-singlet ($\rho_s$)
vector resonance, and for SM-like kinematics of four-top-quark production.

Events are selected with~\cite{atlasss}:
\begin{itemize}
\item Two same-sign leptons (with no veto on additional leptons):
\item Events containing like-flavor leptons are vetoed if $| m_{\ell\ell} - m_Z | \leq 10$~GeV;
\item Missing transverse momentum of at least 40 GeV;
\item Event $H_T$ (scalar sum of lepton and jet transverse momenta) at
  least 350 GeV.
\end{itemize}
The number of selected events and the estimates of the SM
background contributions are shown in Table~\ref{tab:yields}~\cite{atlasss}.

\begin{table}
\center
{\small
\begin{tabular}{lrrr}
\hline\hline
 & $e^\pm e^\pm$ \ \  & $\mu^\pm\mu^\pm$\ \   & $e^\pm\mu^\pm$ \\ \hline \\[-2mm]
\vspace{2mm}
Fake
& $1.0  ^{+0.6}_{-0.7} $ \ \ \ 
& $1.7  ^{+0.7}_{-0.6} $ 
& $3.8  ^{+1.9}_{-1.8} $ 
\\
\vspace{2mm}
Charge flip 
& $0.6 ^{+0.3}_{-0.1} $ \ \ \ 
& $0  ^{+0.1}_{-0.0} $ 
& $0.7   ^{+0.3}_{-0.1} $ 
\\
\vspace{2mm}
Real
& $2.7   ^{+0.7}_{-1.5} $ \ \ \ 
& $2.6   ^{+0.7}_{-0.9} $
& $6.7  ^{+1.3}_{-3.1} $
\\
\hline \\[-2mm]
\vspace{2mm}
Total
& $4.4   ^{+0.5}_{-0.7} $ \ \ \ 
& $4.3  ^{+0.9}_{-1.1} $ 
& $11.2 ^{+2.5}_{-3.6} $
\\ 
\vspace{2mm}
Data 
& 3 \ \ \ 
& 3 
& 12 \\ 
\hline\hline
\end{tabular}
}
\caption{Predicted number of SM background events and observed data  with two
  same-sign leptons, at least two jets, MET $>$ 40~GeV and
  $H_{\textrm T}>350$ GeV, adapted from Ref.~\cite{atlasss}. Uncertainties are
  statistical and systematic in quadrature. }
\label{tab:yields}
\end{table}

In order to interpret these as limits on another process, we first
convert the ATLAS limits on the cross section of a  specific model
into a generic limit on the number of events, $N$, due to a new source.
Using
\[ N = \epsilon \times \sigma \times \mathcal{L}  \]
\noindent
where $\epsilon$ is the fraction of produced events satisfying the
event selection, $\sigma$ is the production cross section and
$\mathcal{L}$ is the integrated luminosity of the data sample.
For $b' \rightarrow tW$ with $m_{b'}=350$ GeV, the ATLAS selection gives  $\epsilon =0.02$ 
and results in an upper limit of $\sigma < 800$ fb at 95\% confidence level~\cite{atlasss}.  
Therefore $N < 16.6$ at 95\% confidence level.  Applying this limit to derive a
cross-section limit for an arbitrary process requires knowing the selection efficiency ($\epsilon$) 
for the model of interest. 

\begin{figure}
\includegraphics[width=0.9\linewidth]{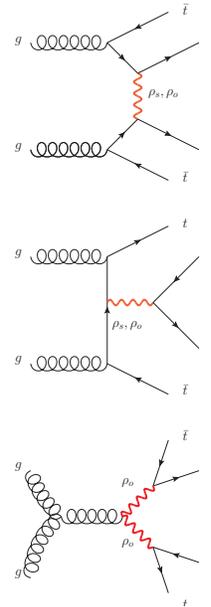}\\
\caption{Representative diagrams of four-top-quark production including the new particles
$\rho_o$ (color octet) and $\rho_s$ (color singlet). In each case, there is a 
representative SM diagram with a gluon in place of a $\rho$.}
\label{fig:diag}
\end{figure}

\begin{figure}
\includegraphics[width=0.98\linewidth]{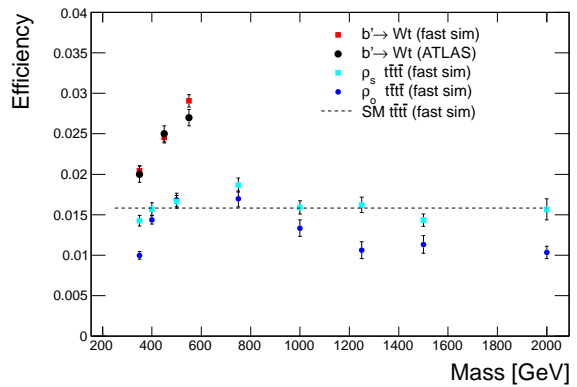}\\
  \caption{ Efficiency of the selection for resonant and SM four-top-quark
    production. For validation, we compare the efficiency using the
    fast simulation to the published ATLAS efficiencies~\cite{atlasss}.}
  \label{fig:eff}
\end{figure}

\begin{figure}
\includegraphics[width=0.7\linewidth]{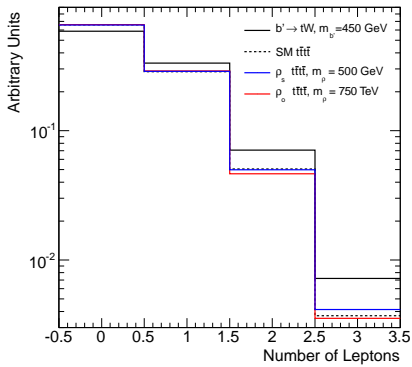}\\
\includegraphics[width=0.7\linewidth]{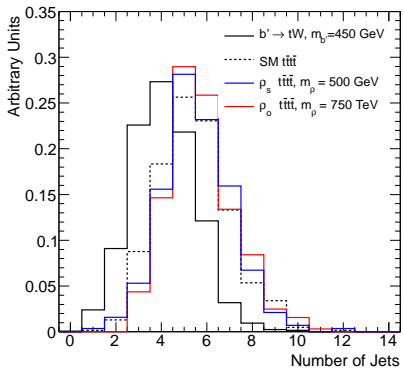}\\
\includegraphics[width=0.7\linewidth]{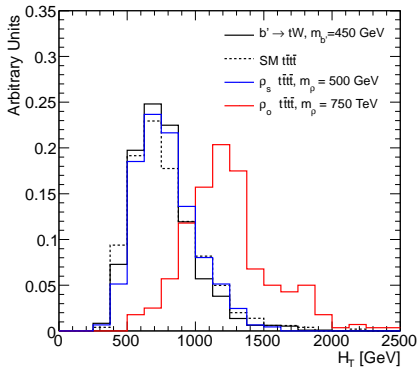}\\
  \caption{ Distribution of kinematic quantities for various
    models. Top, lepton multiplicity before any selection. Center, jet
    multiplicity for events with same-sign dileptons. Bottom, $H_T$
    after all selection requirements other than $H_T>350$.}
  \label{fig:kin}
\end{figure}

We simulate four-top-quark production using {\sc madgraph}~\cite{madgraph}, with {\sc pythia}~\cite{pythia}
for showering and hadronization and detector simulation with a
parametric fast simulation tuned to match ATLAS
performance~\cite{pgs}, for four-top-quark production with SM-like kinematics, and for models
with color-octet or color-singlet vector resonances (see Fig.~\ref{fig:diag}).
The SM prediction for four-top-quark production is around 5~fb \cite{Lillie:2007hd}, well below
the current experimental sensitivity, implying
that any new physics contribution to which the LHC is currently sensitive will
have negligible interference with SM four-top-quark production processes.
Efficiencies for SM-like four-top-quark as well as
color-singlet $\rho_s$ or color-octet $\rho_o$ are shown in
Fig~\ref{fig:eff}.  
Some representative kinematic distributions are shown in Fig.~\ref{fig:kin}.
To validate our efficiency calculation, we compare
published ATLAS efficiencies for $b'\rightarrow tW$ to efficiencies we
measure using the identical production, showering, hadronization and
detector simulation as described above; we find good agreement between
our efficiencies and the published ATLAS efficiencies.
The efficiency for  models with the new $\rho$
particle rises with $m_\rho$ for
low-mass due to increases in the lepton and jet momenta; at high mass,
it falls due to low lepton isolation efficiencies from the increased
activity and the greater top-quark and $W$ boson boosts.

\begin{figure}
\includegraphics[width=0.98\linewidth]{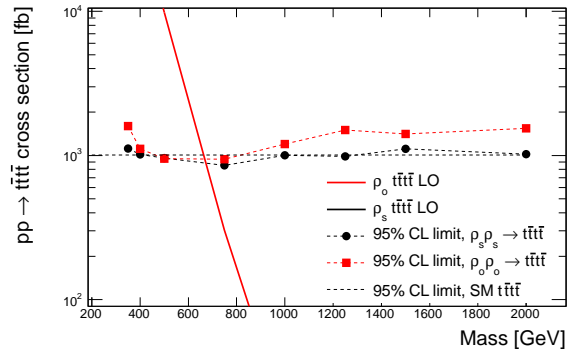}\\
  \caption{ Upper limits at 95\% confidence level on production of a
    four-top-quark final state via SM or $\rho_s,\rho_o$ production. Also
    shown is leading-order predictions in the color-octet model; the
    prediction for SM four-top-quark or color-singlet models are less than
    10 fb.}
  \label{fig:lim}
\end{figure}

Limits are computed from $\sigma < N/(\epsilon \times \mathcal{L})$, 
and are shown in Fig~\ref{fig:lim}.  The limits require a cross section
for anomalous sources of four-top-quark production to be less than about 1~pb, and are
similar for both models containing a $\rho_o$ or $\rho_s$, as well as
models which produce the four top-quarks with SM-like kinematics.  
A re-analysis of CMS dilepton and
trilepton data was performed~\cite{jaas} in a similar spirit
and leads to comparable bounds.
For moderately strongly
coupled $\rho_o$ ($g_{\rho t \bar{t}} \sim 1$), the bound on the $\rho_o$ mass
is around 700 GeV, significantly improving upon the existing Tevatron limits.
Bounds on a four-top-quark contact interaction are currently weak enough so as
to invalidate any hope that the effective theory is a good description,
but should become interesting with more data at higher center-of-mass
energy \cite{Pomarol:2008bh,Kumar:2009vs}.

In closing, we have set bounds on the rate of four-top-quark production, using the recent ATLAS
same-sign dilepton search.  While this search is effective in placing interesting limits on four-top-quark
production, it was optimized for events containing two top quarks, and so one can imagine improving
upon it by using the larger set of handles available from the decay
products of four top quarks.     Our
work is only a first step into what should be a fruitful age of multi-top-quark searches at the LHC.

DW and NZ are supported by grants from the Department of Energy
Office of Science and by the Alfred P. Sloan Foundation. TT
is supported in part by NSF Grant PHY-0970171.

\end{document}